\documentclass[11pt]{article}
\usepackage{Blois,epsfig}

\def\be{\begin{equation}}
\def\ee{\end{equation}}
\def\bea{\begin{eqnarray}}
\def\eea{\end{eqnarray}}

\begin{document}
\begin{center}
\Large{\textbf{Hadronic scattering in the Color Glass Condensate}}
\end{center}

\title{}

\author{Raju Venugopalan}

\address{Physics Department, Brookhaven National Laboratory, Upton, NY, 11973, USA}

\maketitle
\abstracts{
Multi-particle production in QCD is dominated by higher twist contributions. The operator product expansion is not 
very effective here because the number of relevant operators grow rapidly with increasing twist. The Color Glass Condensate (CGC) provides a framework in QCD to systematically discuss ``classical" (multiple scattering) and ``quantum" evolution (shadowing) effects in multi-particle production. The apparently insuperable problem of nucleus-nucleus scattering 
in QCD simplifies greatly in the CGC. A few examples are discussed with emphasis on open problems.}

\section{Introduction: the twist expansion and small x physics}

In the Bjorken limit of QCD, $Q^2\rightarrow \infty$, $s\rightarrow \infty$, $x_{\rm Bj}\approx Q^2/s ={\rm fixed}$, we have a powerful framework to compute a large number of processes to high accuracy. 
Underlying this machinery is the Operator Product Expansion (OPE), where  cross-sections are identified as 
a convolution of short distance "coefficient functions" which are process dependent and long distance 
parton distribution functions which are universal. The evolution of the parton distribution functions with $x$ and $Q^2$ is described by splitting functions, which determine the probability of ``parent" partons 
to split into a pair of ``daughter" partons. Both coefficient functions and splitting functions for DIS inclusive cross-sections are now available to Next-Next-Leading-Order (NNLO) accuracy~\cite{MVV}.  

The leading contributions in the OPE come from operators which have minimal twist, where twist is defined as the 
dimension minus the spin of the operators. Higher twist operators are suppressed by powers of $(m^2/Q^2)^{T-T_{\rm min}}$, 
where $m$ is a hadronic mass scale, $T$ denotes the twist of the operator and $T_{\rm min}$ the minimal twist ($2$ for the structure functions $F_1$ 
and $F_2$). These higher twist operators can therefore be ignored in the Bjorken limit, albeit their contribution provides 
a systematic error to the application of the leading twist formalism at finite $Q^2$. 

However, the bulk of multi-particle scattering  in QCD is dominated by soft and semi-hard physics. In the language of the 
OPE, the latter are higher twist effects. These are of two sorts. The first are ``kinematic" higher twist contributions, which 
arise from trace contributions that are often ignored in the OPE, where the leading contributions are from symmetric 
and {\it traceless} operators~\cite{Nachtmann,Gottleib}. These kinematic high twist contributions are of order $x_{\rm Bj}^2 
m^2/Q^2$ and are of decreasing importance at small $x_{\rm Bj}$.  The other ``dynamical" higher twist contributions are 
from the hadronic matrix elements of the higher twist operators themselves. The relevant twist four matrix elements for leptoproduction were discussed in great detail by Ellis, Furmanski and Petronzio~\cite{EFP} and by Jaffe and Soldate~\cite{JS}, and expressed in terms of multi-parton distributions by Jaffe~\cite{Jaffe}. Discussions specific to twist four 
contributions at small x are contained in Ref.~\cite{Blumlein,Bartels}. There are many more 
contributions at twists greater than four-these have not been quantified. 

To understand why the twist expansion is likely not a useful expansion at small x, we need to consider the properties of the theory in the Regge limit:$x_{\rm Bj}\rightarrow 0$, $s\rightarrow \infty$, $Q^2={\rm fixed}$.  The  BFKL renormalization group equation~\cite{BFKL} describes the leading $\alpha_S \ln(1/x)$ behavior of gluon distributions in this limit. The solutions of the BFKL equation predict that gluon distributions grow 
very rapidly with decreasing $x$.  In the Regge asymptotics, since the transverse size of the partons is 
fixed, this growth of distributions will lead to the overlapping of partons in the transverse plane of the 
hadron. In this regime, contributions that were power suppressed in the BFKL scheme become important. These are recombination and screening effects which slow down the growth of gluon distributions leading ultimately to a saturation of these distributions~\cite{GLR,MuellerQiu}. Such 
effects must appear at small $x$ because the occupation number~\footnote{This corresponds to the number of partons per unit transverse area, 
per unit transverse momentum, in light cone gauge. This condition has its gauge invariant counterpart in the requirement that the field strength 
squared not exceed $1/\alpha_S$.} of partons in QCD can be at most of order $1/\alpha_S$.  

Qualitatively, the competition between Bremsstrahlung 
and recombination/screening effects become of the same order when 
${1\over 2\,(N_c^2 -1)}\,{x\,G(x,Q^2)\over \pi R^2 Q^2} \approx {1\over \alpha_S(Q^2)}$.
This relation is solved self-consistently when $Q\equiv Q_s(x)$. The scale $Q_s(x)$ is termed the saturation scale and it grows as one goes to smaller values of $x$. What does this have to do with higher twists? As we will discuss further in 
the next section, when $Q_s(x)^2 \geq Q^2$, all higher twists are equally important. The OPE therefore is not a good 
expansion in this small x kinematic region, where the typical momentum of partons is of order $Q_s$~\footnote{Another more 
severe reason why the OPE breaks down-even at leading twist at small 
x-has to do with the infrared diffusion in the BFKL equation~\cite{Mueller_OPE}. Ironically, this diffusion is cured by 
higher twist saturation effects.}. 

There is however a glimmer of hope in this seemingly hopeless situation. This is because $Q_s^2(x) >> \Lambda_{\rm QCD}^2$, which suggests that weak coupling techniques in QCD are applicable in the Regge limit. In the next section, we will discuss a weak coupling effective field theory approach which provides a more efficient organizing principle than the OPE at high parton densities.

\section{The Color Glass Condensate}

The physics of high parton densities can be 
formulated as a classical effective theory~\cite{MV,IV} because there is a Born-Oppenheimer separation between large x and small x 
modes~\cite{Susskind} which are respectively the slow and fast modes in the effective theory.  Large x partons are static sources of color charge for the dynamical wee (small x) parton fields. The generating functional 
of wee partons has the form
\begin{eqnarray}
{\cal Z}[j] = \int [d\rho]\, W_{\Lambda^+}[\rho]\,\left\{ {\int^{\Lambda^+} [dA] \delta(A^+) e^{iS[A,\rho]-j\cdot A} \over \int^{\Lambda^+} [dA] \delta(A^+) e^{iS[A,\rho]}}\right\} 
\label{eq:2}
\end{eqnarray}
where the wee parton action has the form
\begin{eqnarray}
S[A,\rho] ={-1\over 4}\,\int d^4 x \,F_{\mu\nu}^2
+ {i\over N_c}\, \int d^2 x_\perp dx^- \delta(x^-)\,{\rm Tr}\left(\rho(x_\perp)U_{-\infty,\infty}[A^-]\right) \, .
\label{eq:3}
\end{eqnarray}
In Eq.~\ref{eq:2}, $\rho$ is a two dimensional classical color charge density and 
$W[\rho]$ is a weight functional of sources (which sit at momenta $k^+ > \Lambda^+$: note, 
$x = k^+/P_{\rm hadron}^+$). The sources are coupled to the dynamical wee gluon fields (which in 
turn sit at $k^+ < \Lambda^+$) via the gauge invariant term~\footnote{For an alternative gauge invariant form, which also 
recovers the BFKL equation, see Ref.~\cite{SJR}.} which is the second term on the RHS of 
Eq.~\ref{eq:3}.  Here $U_{-\infty,\infty}$ denotes a path ordered exponential of the gauge field $A^-$ in the $x^+$ direction. The first term in Eq.~\ref{eq:3} is the QCD field strength tensor squared-thus the wee 
gluons are treated in full generality in this effective theory, which is formulated in the light cone gauge $A^+=0$. The source $j$ is an external source-derivatives taken with respect to this source (with the source then put to zero) generate correlation functions in the usual fashion. 

The argument for why the sources are classical is subtle and follows from a coarse graining of the effective action. The weight functional for a large nucleus is a Gaussian in the source density ~\cite{MV,Kovchegov}, with a small correction for SU($N_c$) coming from the $N_c-2$ higher Casimir operators~\cite{SR1}.  
The variance of the Gaussian, the color charge squared per unit area $\mu_A^2$, 
proportional to $A^{1/3}$, is a large scale-and is the only scale in the effective action~\footnote{$\mu_A^2$ is simply related in the classical 
theory to the saturation scale $Q_s^2$ via the relation $Q_s^2 =\alpha_S N_c \mu_A^2 \ln(Q_s^2/\Lambda_{\rm QCD}^2)$}. Thus for $\mu_A^2 >> \Lambda_{\rm QCD}^2$, $\alpha_S(\mu_A^2) <<1$, and one can compute the properties of the theory in Eq.~\ref{eq:2} in weak coupling. For an SU(3) Yang-Mills theory, there is an additional contribution proportional to the 
cubic Casimir operator. It is parametrically suppressed by $A^{1/6}$. However, this term generates Odderon excitations in the 
CGC already at the classical level~\cite{SR2}. 

The saddle point of the action in Eq.~\ref{eq:3} gives the classical distribution of gluons in the nucleus. The Yang-Mills 
equations can be solved analytically to obtain the classical field of the nucleus as a function of $\rho$: $A_{\rm cl.}(\rho)$~\cite{MV,Kovchegov,JKMW}. One can determine, for Gaussian sources,  the occupation number $\phi = dN/\pi R^2/ dk_\perp^2 dy$ of wee partons in the classical field of the nucleus.
One finds for $k_\perp >> Q_s^2$, the Weizs\"acker-Williams spectrum $\phi \sim Q_s^2 / k_\perp^2$; for $k_\perp \leq Q_s$, one obtains a complete resummation to all 
orders in $k_\perp$, which gives $\phi\sim {1\over \alpha_S} \ln(Q_s/k_\perp)$. (The behavior at low $k_\perp$ can, more accurately, be represented 
as ${1\over \alpha_S} \Gamma(0,z)$ where $\Gamma$ is the incomplete Gamma function and $z = k_\perp^2/Q_s^2$~\cite{Dionysis}). 

Small fluctuations about the effective action in Eq.~\ref{eq:3} give large corrections~\cite{AJMV} of order $\alpha_S\ln(1/x)$. The Gaussian weight functional is thus fragile under quantum evolution of the sources.  
A Wilsonian renormalization group (RG) approach systematically 
treats these corrections~\cite{JIMWLK}.  In particular, the change of the weight functional $W[\rho]$ with x is described by 
the JIMWLK- non-linear RG equations~\cite{JIMWLK}.
These equations form an infinite hierarchy of 
ordinary differential equations for the gluon correlators $<A_1 A_2 \cdots A_n>_Y$, where $Y= \ln(1/x)$ is the rapidity. For the gluon density, which is proportional to a two-point function 
$<\alpha^a (x_\perp) \alpha^b (y_\perp)>$, one recovers the BFKL equation in the limit of low parton densities. Further 
developments beyond the JIMWLK equation have been summarized at this conference by Iancu~\cite{Iancu}. 

In the limit of large $N_c$ and large A ($\alpha_S^2 A^{1/3} >> 1$), the JIMWLK hierarchy closes for the two point correlator of Wilson lines because the expectation value of the product of traces of Wilson lines factorizes into the product of the expectation values of the traces:
$\langle{\rm Tr}(V_x V_z^\dagger) {\rm Tr}(V_z V_y^\dagger)\rangle \longrightarrow \langle{\rm Tr}(V_x V_z^\dagger)\rangle\,\langle{\rm Tr}(V_z V_y^\dagger)\rangle$,
where $V_x={\cal P}\exp\left(\int dz^- \alpha^a(z^-,x_\perp) T^a\right)$. Here ${\cal P}$ denotes path ordering in $x^-$ and $T^a$ is an adjoint SU(3) 
generator. The cross-section for a $q{\bar q}$ pair 
scattering off a target can be expressed in terms of these 2-point dipole operators as
$\sigma_{q\bar q N} (x, r_\perp) = 2\, \int d^2 b\, \, {\cal N}_Y (x,r_\perp,b)$, 
where ${\cal N}_Y$, the imaginary part of the forward scattering amplitude, is defined to be ${\cal N}_Y= 1 - {1\over N_c}<{\rm Tr}(V_x V_y^\dagger)>_Y$. 
The size of the dipole, ${\vec r}_\perp = {\vec x}_\perp - {\vec y}_\perp$ and ${\vec b} = ({\vec x}_\perp + {\vec y}_\perp)/2$. The JIMWLK equation for the two point Wilson correlator is identical in the 
large A, large $N_c$ mean field limit to an equation derived independently by Balitsky and Kovchegov-the Balitsky-Kovchegov equation~\cite{BK}, which has the operator form
${{\partial {\cal N}_Y}\over \partial Y} = {\bar \alpha_S}\, {\cal K}_{\rm BFKL} \otimes \left\{ {\cal N}_Y - {\cal N}_Y^2\right\}$.
Here ${\cal K}_{\rm BFKL}$ is the well known BFKL kernel. When ${\cal N} << 1$, the quadratic term is negligble and one has BFKL growth of the number of dipoles; when ${\cal N}$ is close to unity, 
the growth saturates. The B-K equation is the simplest equation including
both the Bremsstrahlung responsible for the rapid growth of amplitudes at small x as well as the repulsive many 
body effects that lead to a saturation of this growth. 

We now return to our discussion of higher twists in the previous section. In lepto-production, the structure function 
$F_2$ at small x is proportional to $|\psi_{\gamma^\star\rightarrow q{\bar q}}|^2\,\otimes \sigma_{q{\bar q} N}$, where 
$|\psi_{\gamma^\star\rightarrow q{\bar q}}|^2$ is the probability for a virtual photon to split into a $q{\bar q}$ pair and $\sigma_{q{\bar q} N}$ is the $q{\bar q}-N$ cross-section discussed previously. Since the latter is proportional to a product of 
Wilson lines, $F_2$ gets contributions from $N$-point gluon distributions. In the classical McLerran-Venugopalan (MV) model of Gaussian color sources, these 
can be expressed explicitly~\cite{MV99} as an expansion in $Q_s^2/Q^2$-thus, for $Q_s^2 \geq Q^2$, all higher twists contribute equally. 
The OPE would not be very useful in this region-however, in the CGC framework, higher twist effects are included both 
at the tree level in the MV model, and in the small x quantum evolution of the BK and JIMWLK RG equations. It is interesting 
to ask whether the effective theory of the CGC at sufficiently small x and large $Q^2$ can be matched on to the full theory 
beyond the leading twist level-this has not been done thus far but is feasible in principle.

\section{Hadronic Scattering in the CGC}

At small x, both the collinear factorization and $k_\perp$ factorization limits of pQCD can be understood in a systematic way in the framework of the CGC. Rather than a convolution of probabilities, one has instead the collision of classical gauge fields. The expectation value of an operator ${\cal O}$ can be computed as 
\begin{eqnarray}
<{\cal O}>_Y = \int [d\rho_1]\, [d\rho_2]\, W_{x_1}[\rho_1]\, W_{x_2}[\rho_2]\, {\cal O}(\rho_1,\rho_2) \, ,
\label{eq:9}
\end{eqnarray}
where $Y = \ln(1/x_F)$ and $x_F = x_1 - x_2$. All operators at small $x$ can be computed in the background 
classical field of the nucleus at small $x$. Quantum 
information, to leading logarithms in $x$, is contained in the source functionals $W_{x_1 (x_2)}[\rho_1(\rho_2)]$. The operator 
${\cal O}$ can be expressed in terms of gauge fields $A^\mu[\rho_1,\rho_2](x)$. 

Inclusive gluon production in the CGC is computed by solving the Yang-Mills equations 
$[D_\mu,F^{\mu\nu}]^a = J^{\nu,a}$, where 
$J^\nu = \rho_{1}\,\delta(x^-)\delta^{\nu +} + \rho_{2}\,\delta(x^+)\delta^{\nu -}$, 
with initial conditions given by the Yang-Mills fields of the two nuclei before the collision. These are obtained 
self-consistently by matching the solutions of the Yang-Mills equations on the light cone~\cite{KMW}. The initial conditions 
are determined by requiring that singular terms in the matching vanish. Since the Yang-Mills fields in the nuclei before the collision are known, the classical problem is in principle completely solvable. Quantum corrections not 
enhanced by powers of $\alpha_S\ln(1/x)$ can be included systematically. The terms so enhanced are absorbed into the weight functionals $W[\rho_{1,2}]$.  Thus all ``classical" multiple scattering effects 
are obtained by solving the Yang-Mills equations, while the small x quantum evolution effects (which gives rise to shadowing) are  
contained in the weight functionals which obey the JIMWLK/BK equations. 

Hadronic scattering in the CGC can therefore be studied through a systematic power counting in the density of sources in powers of $\rho_{1,2}/k_{\perp;1,2}^2$. 
The power counting is applicable either to a proton at small x, or to a nucleus (whose parton density 
at high energies is enhanced by $A^{1/3}$) at large transverse momenta. The 
relevant quantity here is $Q_s$, which, as one may recall, is enhanced both for large $A$ and small $x$. 
So as long as $k_\perp >> Q_s >> \Lambda_{\rm QCD}$, the proton or nucleus is considered dilute. 
One can begin to study the applicability of both collinear and $k_\perp$ factorization at small $x$ in this 
approach. 

To lowest order in $\rho_{p1}/k_\perp^2$ and $\rho_{p2}/k_\perp^2$, one can compute inclusive gluon production analytically~\cite{KMW}. At large transverse momenta, $Q_s << k_\perp$, the scattering can be expressed in a $k_\perp$-factorized 
 form. The inclusive cross-section is expressed as the product of 
 two unintegrated ($k_\perp$ dependent) distributions times the matrix element for the scattering. 
 $k_\perp$ factorization is a good assumption at large momenta for quark pair-production. This was worked out in the CGC approach by Fran\c cois Gelis and myself~\cite{FR}.  In this limit, our result agrees exactly with that of Refs.~\cite{CCHCE}.

In the semi-dense/pA case, one solves the Yang--Mills equations  
to determine the gluon field produced-to lowest order in the proton source density and to all orders in the  
nuclear source density.  The inclusive gluon production cross-section, in this framework, was first computed in Refs.~\cite{KovMueller,DumitruMcLerran} and shown to be $k_\perp$ factorizable in Ref.~\cite{KKT}. In Ref.~\cite{BFR1}, the gluon field produced in pA collisions was computed explicitly in Lorentz gauge $\partial_\mu A^\mu=0$. The fact that the 
distributions are $k_\perp$ factorizable is remarkable because the ``unintegrated" gluon distribution of the nucleus is not 
the usual leading twist unintegrated distribution, but includes all higher twists. Its evolution with energy is given by the 
JIMWLK/BK equations. 

If we wish to study multiple scattering effects alone, these can be studied in the MV model-which 
provides the initial conditions for quantum evolution. Thus at larger x's multiple scattering effects dominate while quantum effects 
turn on as one goes to smaller x due to the RG evolution of the weight functionals. 
The well known ``Cronin" effect is obtained in our formalism and can be simply understood in terms of the multiple scattering of a parton from the projectile with those in the target. The remarkable energy dependence of the Cronin effect seen by 
the RHIC experiments may be due to quantum evolution effects and has been discussed elsewhere~\cite{RV}.

Quark production in p/D-A collisions can be computed with the gauge field in Lorentz gauge~\cite{BFR2}.  Unlike gluon production, neither quark pair-production nor single quark production is strictly $k_\perp$ factorizable. 
The pair production cross-section however can still be written in $k_\perp$ factorized form as a product of the unintegrated gluon distribution in the proton times a sum of terms with three unintegrated distributions,  $\phi_{g,g}$, $\phi_{q\bar q, g}$ and $\phi_{q\bar q,q\bar q}$. These are respectively proportional to 2-point, 3-point and 4-point correlators of the Wilson lines we discussed previously. For instance, the distribution $\phi_{q\bar q, g}$ is the product of fundamental Wilson lines coupled to a $q\bar q$ pair in the amplitude and adjoint Wilson lines coupled to a gluon in the complex conjugate amplitude. For large transverse momenta 
or large mass pairs, the 3-point and 4-point distributions collapse to the unintegrated gluon distribution, and we recover the previously discussed $k_\perp$-factorized result for pair production in the dilute/pp-limit. Single quark distributions are straightforwardly obtained and depend only on 
the 2-point quark and gluon correlators and the 3-point correlators-which are ``all twist" operators as previously. For Gaussian sources, as in the MV-model, these 2-,3- and 4-point functions can be computed exactly as discussed in Ref.~\cite{BFR2}. The renormalization group evolution of these distributions lead to shadowing of the distributions. Understanding their evolution with 
energy may provide important information about the structure of multi-parton correlations in high energy QCD. 

Our results for gluon and quark production in p/D-A collisions (for a review, see Ref.~\cite{YuriJamal}), coupled with the previous results for 
inclusive and diffractive~\cite{GelisJamal} distributions in DIS suggest an important new paradigm. At small $x$ in DIS and hadron colliders, quark and gluon structure functions, which are the right 
observables in a leading twist formalism, are 
no longer the right observables to capture the relevant physics. Instead they should be replaced by  dipole and 
multipole correlators of Wilson lines that seem ubiquitous in all high energy processes and are similarly gauge invariant~\footnote{This relies on 
requiring that one closes the ends of the Wilson lines at $\pm \infty$ in the nucleus to form a closed loop, and further that the fields go to zero 
at $\pm \infty$. While physically attractive, this construction is arbitrary. I thank F. Gelis and K. Kajantie for a discussion on this point.}
and process independent. For a similar conclusion on the importance of unintegrated distributions from a different perspective, see Ref.~\cite{CollinsJung}. To determine whether these distributions are robust, next-to-leading order 
computations must be performed, which is a formidable, but by no means impossible, task.

In nucleus-nucleus collisions, $\rho_{1,2}/k_\perp^2\sim 1$. There is no small expansion parameter and one has thus far not been able to compute particle production analytically in the CGC.  
Unlike gluon production in the pp and pA cases, $k_\perp$-factorization breaks down in the AA-case~\cite{AR,Balitsky2}.  A significant consequence is that one cannot factor the quantum evolution of the initial 
wavefunctions into unintegrated gluon distributions unlike the pA case.
Nevertheless, there is a systematic way to 
include small x effects in the AA case. The problem of nuclear collisions is well defined in weak coupling and can be solved numerically~\cite{AR}.
The numerical simulations thus far assume Gaussian distributions of the color sources as in the MV model. This is 
reasonable for central Gold-Gold collisions at RHIC where the typical $x$ is of order $10^{-2}$. At the LHC, the 
typical $x$ at central rapidities is an order of magnitude lower. At these x values, quantum evolution effects are important 
and one should use one solutions of JIMWLK/BK RG equations to determine the distribution of sources.

We  will restrict ourselves to discussing numerical solutions for  Gaussian color sources. The saturation 
scale $Q_s$ (which is an input in the numerical solutions in this approximation) and the nuclear radius $R$ are the 
only parameters in the problem. 
The energy and number respectively of gluons released in a heavy ion collision of identical nuclei can therefore be simply expressed as ${1\over \pi R^2}\,{dE\over d\eta} =] {c_E\over g^2}\,Q_s^3$ and 
${1\over \pi R^2}\,{dN\over d\eta} = {c_N\over g^2}\, Q_s^2$, 
where (up to $10\%$ statistical uncertainity) we compute numerically~\cite{AR} $c_E=0.25$ and $c_N=0.3$. Here $\eta$ is the space-time rapidity. 
The number distributions of gluons can also be computed in this approach. Remarkably, one finds that a) the 
number distribution is infrared finite, and b) the distribution 
is well fit by a massive Bose-Einstein distribution for $k_\perp/Q_s < 1.5$ GeV with a ``temperature" of $\sim 0.47 Q_s$ 
and by the perturbative distribution $Q_s^4/k_\perp^4$ for $k_\perp/Q_s > 1.5$. We will not discuss comparisons of CGC 
predictions to RHIC data here but refer to Ref.~\cite{KLN}. 

The transition to the QGP from the CGC remains as an outstanding theoretical problem.  Due to the rapid expansion of the system, the occupation number of modes falls well below one on time scales of order $1/Q_s$. From these times onwards, the classical approach breaks down for all but the softest modes-well before thermalization.  On the other hand, for elliptic flow from hydrodynamics to be significant, the conventional wisdom is that thermalization should set in early. A necessary condition is that momentum distributions should be isotropic. The CGC initial conditions are very anisotropic with $<p_\perp> \sim Q_s$ and $<p_z>\sim 0$. How does this isotropization take place? All estimates of final state re-scattering of partons formed from the melting CGC, both from $2\rightarrow 2$ 
processes~\cite{Mueller6} and $2\rightarrow 3$ processes~\cite{Wong} suggest thermalization takes 
longer than what the RHIC collisions seem to suggest. In the ``bottom up" scenario~\cite{BMSS}, $\tau_{\rm thermal}\sim {1\over \alpha_S^{13/5}}\,{1\over Q_s}$, which at RHIC energies gives $\tau_{\rm thermal}\sim 2-3$ fm. 

Recently, it has been suggested that collective instabilities, analogous  to the well known Weibel instabilities 
in plasma physics, can speed up themalization~\cite{AML}. For a nice recent review, see Ref.~\cite{Stan}. Starting from very anisotropic (CGC-like) initial conditions, these instabilities drive the system to isotropy. In some estimates, these may 
be very short time scales of order $1/Q_s$. 
What is the relation of this language of instabilities and that of our classical field simulations? Paul Romatschke and I~\cite{PaulRaju} recently looked at the effect  of small violations of boost invariance on the dynamics of gauge fields produced with CGC initial conditions. We found that there is a Weibel instability and that the maximally unstable modes 
grow as $\exp(\sqrt{\tau})$ (as opposed to $\exp(\tau)$ in a static box). This behavior was predicted in Ref.~\cite{ALM}. 
Increasing the initial amplitude of the fields that violate boost invariance, we find that the growth of the instability terminates 
when non-Abelian effects become large. These studies don't address the possible isotropization of the system by instabilities 
because the initial amplitudes examined are much smaller than physically plausible. This work is in progress. 

An equally interesting problem is that of chemical equilibration.  At high energies, the initial state in a heavy ion collision is dominated by gluons. Are quarks produced in sufficient numbers for the system to reach chemical equilibrium (where the ratio of gluons to quarks is expected to be $32/21 N_f$)? Recently, the Dirac equation for quarks was solved by Gelis, Kajantie 
and Lappi~\cite{GKL} in the background field computed in Ref.~\cite{AR}. For $\alpha_S$ values comparable to those in the pre-equilibrium phase of 
RHIC collisions, they indeed find that sufficient numbers of quarks pairs are produced for chemical equilibrium to be plausible. 
As for the studies of collective instabilities, more intensive numerical studies are required for conclusive results.

\section*{Acknowledgments}

This work was supported by DOE Contract No. DE-AC02-98CH10886 . I would like to thank Basarab Nicolescu and his 
colleagues for a very enjoyable meeting.

\end{document}